\documentclass[sigconf]{acmart}
\AtBeginDocument{\DeclareCaptionSubType{lstlisting}}

\usepackage[all]{nowidow}
\usepackage[english]{babel}
\usepackage[strict]{changepage}
\usepackage[T1]{fontenc}
\usepackage[utf8]{inputenc} 
\usepackage{amsmath,amsfonts}
\usepackage{amsthm}
\usepackage{anyfontsize}
\usepackage{array}
\usepackage{booktabs}
\usepackage{color}
\usepackage{comment}
\usepackage{etoolbox}
\usepackage{float}
\usepackage{graphicx}
\usepackage{hyperref}
\usepackage{hyphenat}
\usepackage{lineno}
\usepackage{listings}
\usepackage{makecell}
\usepackage{mdframed}
\usepackage{multirow}
\usepackage{setspace}
\usepackage{stfloats}
\usepackage{subcaption}
\usepackage{syntax}
\usepackage{tcolorbox}
\usepackage{textcomp}
\usepackage{url}
\usepackage{verbatim}
\usepackage{xspace}
\usepackage{colortbl}
\usepackage{array}
\usepackage{multirow}
\usepackage{ragged2e}
\usepackage{colortbl}
\usepackage{hhline}
\usepackage[linesnumbered]{algorithm2e}

\hypersetup{
    colorlinks=true,
    linkcolor=blue,
    filecolor=magenta,      
    urlcolor=black,
    citecolor=violet,
    pdfpagemode=FullScreen,
    }

\definecolor{light-gray}{gray}{0.85}

\mdfsetup{
  skipabove=0.5em,
  skipbelow=0.5em,
  innertopmargin=0.5em,
  innerbottommargin=0.5em
}

\let\oldnl\nl
\newcommand{\nonl}{\renewcommand{\nl}{\let\nl\oldnl}}

\newcommand{\review}[1]{{\color{black}#1\normalfont}}
\newcommand{\pysr}{\textsc{SafeRefactorPy}}

\newcommand{\totalFailures}{480}

\newcommand{\totalCannotApply}{197}

\newcommand{\totalCorrect}{354}

\newcommand{\totalTargets}{\review{1,152}}
\newcommand{\totalBugs}{29}

\newcommand{\totalDistinctTypeErrors}{18}

\newcommand{\totalKeywords}{1}
\newcommand{\totalFields}{12}
\newcommand{\totalMethods}{18}

\newcommand{\code}[1]{\texttt{#1}}
\newcommand{\typeError}[1]{\textit{#1}}

\usepackage{tikz}

\microtypesetup{activate={true,nocompatibility}, patch=none}

\setcopyright{none}
\renewcommand\footnotetextcopyrightpermission[1]{}

\begin{document}

\title{Bugs in the Shadows: Static Detection of Faulty Python Refactorings}

 \author{Jonhnanthan Oliveira}
 \orcid{0000-0002-7782-410X}
 \affiliation{%
   \normalsize \institution{Federal University of Campina Grande} \country{Brazil}
 }
 \email{jonhnanthan@copin.ufcg.edu.br}

 \author{Rohit Gheyi}
 \orcid{0000-0002-5562-4449}
 \affiliation{%
   \normalsize \institution{Federal University of Campina Grande} \country{Brazil}
 }
 \email{rohit@dsc.ufcg.edu.br}
 
 \author{Márcio Ribeiro}
 \orcid{0000-0002-4293-4261}
 \affiliation{%
   \normalsize \institution{Federal University of Alagoas} \country{Brazil}
 }
 \email{marcio@ic.ufal.br}

 \author{Alessandro Garcia}
 \orcid{0000-0001-5788-5215}
 \affiliation{%
   \normalsize \institution{Pontifical Catholic University of Rio de Janeiro} \country{Brazil}
 }
 \email{afgarcia@inf.puc-rio.br}

\begin{abstract}
Python is a widely adopted programming language, valued for its simplicity and flexibility. 
However, its dynamic type system poses significant challenges for automated refactoring -- an essential practice in software evolution aimed at improving internal code structure without changing external behavior.
Understanding how type errors are introduced during refactoring is crucial, as such errors can compromise software reliability and reduce developer productivity.
In this work, we propose a static analysis technique to detect type errors introduced by refactoring implementations for Python. 
We evaluated our technique on Rope refactoring implementations, applying them to open-source Python projects.
Our analysis uncovered \totalBugs{} bugs across four refactoring types from a total of \totalTargets{} refactoring attempts.
Several of these issues were also found in widely used IDEs such as PyCharm and PyDev.
All reported bugs were submitted to the respective developers, and \review{some of them} were acknowledged and accepted.
These results highlight the need to improve the robustness of current Python refactoring tools to ensure the correctness of automated code transformations and support reliable software maintenance.
\end{abstract}


\maketitle

\section{Introduction}

Python is a widely used programming language, valued for its simplicity and flexibility~\cite{Rossum-book-2011}. 
Its syntax supports multiple paradigms -- such as imperative, object-oriented, and functional -- and allows developers to choose between typed and untyped code. 
While these features contribute to Python's popularity, they also complicate code maintenance and tooling support, especially for automated refactoring~\cite{Fowler-book-1999,Opdyke-PHD-1992,Mens-TSE-2004}.

Refactoring is a key practice in software evolution that aims to improve internal code structure without changing its external behavior. 
Despite its conceptual elegance, implementing correct and reliable refactoring transformations is non-trivial~\cite{Tempero-ACM-2017,Schafer-OOPSLA-2010,tip-oopsla-2003,schafer-ecoop-2009,schafer-oopsla-2008,Steimann-ecoop-2009}. 
This task becomes even more challenging in dynamically typed languages such as Python~\cite{Schafer-fwrt-2012}, where type-related issues can easily escape compile-time checks and manifest at runtime as errors or unintended behavior.

\review{Prior research has introduced techniques for evaluating the correctness of refactoring tools in languages other than Python.} 
For instance, Sch\"{a}fer et al.~\cite{Schafer-OOPSLA-2010} developed an intermediate representation to simplify the specification and verification of \review{Java} refactorings. 
Gligoric et al.~\cite{Gligoric-ecoop-13} tested refactoring implementations in Eclipse for Java and C, uncovering 120 bugs, nearly 30\% of which were type-related. 
Other studies, such as those by Mongiovi et al.~\cite{Mongiovi-icsme-2014,Mongiovi-TSE-2018} and Soares et al.~\cite{Soares-TSE-2013,Soares-ICSM-2011}, investigated the presence of compilation errors, behavioral changes, and overly strong preconditions in Java-based refactoring engines.
In contrast, the extent to which Python refactoring tools introduce type errors remains underexplored. 
Given the increasing reliance on Integrated Development Environments (IDEs) that support automated refactoring, ensuring the robustness of these tools is crucial. 
\review{There is an increasing demand for improved tooling support among Python developers~\cite{golubev-fse-2021,Wang-icsme-2025}.}
Tools must handle Python's dynamic typing without introducing type errors, behavioral regressions, or incorrect precondition assumptions~\cite{sun-ase-2022,Mongiovi-TSE-2018,Soares-TSE-2013,Soares-ICSM-2011,Oliveira-2019-IST,Schafer-PLPV-2009,Pinto-wrt-2013}.

In this paper, we propose a technique \review{in Section~\ref{sec:technique}} to statically detect type errors introduced by refactoring implementations in Python (\review{referred to as \pysr{}}). 
Our approach applies each refactoring transformation individually to real-world Python projects and uses a differential analysis technique based on Pyre~\cite{pyre}, a static type checker for Python, to compare the type-checking results between the original and refactored versions of the code. 
We evaluate \review{in Section~\ref{sec:evaluation}} our technique by applying \totalTargets{} transformations from refactorings implemented in Rope~\cite{rope}. 
Our technique identified \totalBugs{} bugs across four refactoring types. 
Manual evaluation of Rope's transformations in popular IDEs such as PyCharm, PyDev, and VSCode revealed that some of these bugs also appear in PyCharm and PyDev.
We reported all discovered bugs to the relevant developers, and \review{some of them} were acknowledged and accepted.

The results of our study have important implications for the development and adoption of automated refactoring tools in dynamically typed languages such as Python. Our findings reveal that even widely used refactoring engines can introduce subtle type-related bugs, which may go unnoticed by developers and result in runtime \review{errors} or degraded software quality. This highlights the need for rigorous validation mechanisms in refactoring tools to ensure semantic preservation and correctness of transformations. By statically detecting type errors introduced by refactorings, our approach can serve as a complementary validation step during tool development or integration in IDEs, ultimately contributing to more reliable code evolution workflows. Furthermore, the acceptance of our bug reports by tool maintainers underscores the practical relevance and real-world applicability of our technique.
In summary, our study contributes the following:
\begin{itemize}
    \item We introduce a static analysis technique to identify type errors in Python refactoring implementations (Section~\ref{sec:technique});
    \item We apply the technique to \totalTargets{} transformations, identifying \totalBugs{} bugs in 4 refactoring implementations (Section~\ref{sec:evaluation}).
\end{itemize}
All study artifacts are available online~\cite{artifacts}.

\section{Motivating Example}
\label{sec:motivatingExample}

This section presents a motivating example of a type error introduced by a refactoring in Python, illustrating the kinds of challenges our study seeks to address.
Consider a Python program that reads data from a CSV file and compares instances based on an initial value. 
Listing~\ref{lstMotivating:inputProgram} shows a Python class, \code{Mark}, which defines \review{function}s for setting the initial state\footnote{\url{https://docs.python.org/3/reference/datamodel.html\#object.__init__}} and comparing two objects. 
This example is adapted from the source code of the \textit{TextBlob} project.

\begin{lstlisting}[language=Python, caption={Initial Python Program.}, label=lstMotivating:inputProgram]
import csv
class Mark(object):
    def __init__(self, marks, fp):
        self._marks = marks
        reader = csv.reader(
            fp,
            delimiter=';'
        )
        for row in reader:
            print(row)
    def key(self):
        return self._marks
    def %*\colorbox{light-gray}{__lt__}*)(self, other): %*\label{input:target}*)
        return self.key() < other.key()
with open('some.csv') as csvfile:
    mark1 = Mark(9, csvfile)
    mark2 = Mark(8, csvfile)
print(mark1 < mark2) %*\label{input:target2}*)
\end{lstlisting}

Suppose we want to apply the Rename Method refactoring using Rope~\cite{rope} to change the name of the \code{\_\_lt\_\_} \review{function} in Listing~\ref{lstMotivating:inputProgram} to \code{compare}, aiming to better reflect its functionality as a comparator of objects based on key values.
According to its specification~\cite{Fowler-book-1999}, the Rename Method refactoring implementation in Rope updates both the \review{function} declaration and all of its references.
\review{The resulting program replaces all occurrences of \code{\_\_lt\_\_} with \code{compare}.}

However, when executing the program, the Python interpreter\footnote{\url{https://docs.python.org/3/tutorial/interpreter.html\#invoking-the-interpreter}} reads the CSV file and raises a runtime error. 
This occurs because \code{compare} is not a recognized \review{function} in Python's rich comparison model\footnote{\url{https://docs.python.org/3/reference/datamodel.html\#object.__lt__}}, which relies on specific dunder \review{functions} (e.g., \code{\_\_lt\_\_}, \code{\_\_gt\_\_}) for object comparisons.
In this case, the refactoring implementation introduces an error by renaming a special \review{function} that is semantically tied to Python's runtime behavior.
This transformation should have been disallowed by a precondition that detects such special \review{function}s.
This bug is documented in Rope's issue tracker.\footnote{\label{motivateExampleIssue}\url{https://github.com/python-rope/rope/issues/773}}
It highlights the need for refactoring tools to include semantic awareness in their precondition checks, especially when dealing with language-specific conventions and reserved \review{function} names.

Ideally, such type-related errors should be detected statically by the refactoring engine, rather than being discovered only at runtime through execution \review{errors}.
This example illustrates a broader challenge stemming from Python's dynamic type system.
Refactorings involving renaming \review{functions}, fields, or classes may result in runtime errors if all relevant references and semantic constraints are not properly accounted for.
Due to the dynamic and implicit nature of name resolution and \review{function} dispatch in Python, accurately tracking the usage and meaning of identifiers is inherently difficult, especially for automated tools.
Even widely adopted IDEs like PyCharm may fail to capture all contextual interactions, leading to incorrect or incomplete transformations.

Previous work has investigated the correctness of refactoring tools in statically typed languages such as Java and C by detecting compilation errors, behavioral deviations, and overly strong preconditions~\cite{daniel-fse-07,Gligoric-ecoop-13,Soares-TSE-2013,Mongiovi-TSE-2018,Soares-ICSM-2011}.
However, the impact of such refactorings on Python, particularly with respect to type errors, remains largely unexplored.
To address this gap, there is a clear need for techniques capable of statically analyzing refactoring implementations to identify missing preconditions and prevent runtime type errors.
For instance, a robust refactoring tool should verify that renaming a \review{function} like \code{\_\_lt\_\_} to \code{compare} violates Python's comparison protocol before applying such a transformation.
In the following section, we present a technique designed to address this issue by statically detecting type errors introduced during refactoring in Python.

\section{Detecting Type Errors in Refactorings}
\label{sec:technique}

Next we present our technique in detail (\review{\pysr{}}).

\subsection{Overview}
\label{subsec:techOverview}

\begin{figure*}[!ht]
    \centering
    \includegraphics[keepaspectratio, width=0.7\textwidth]{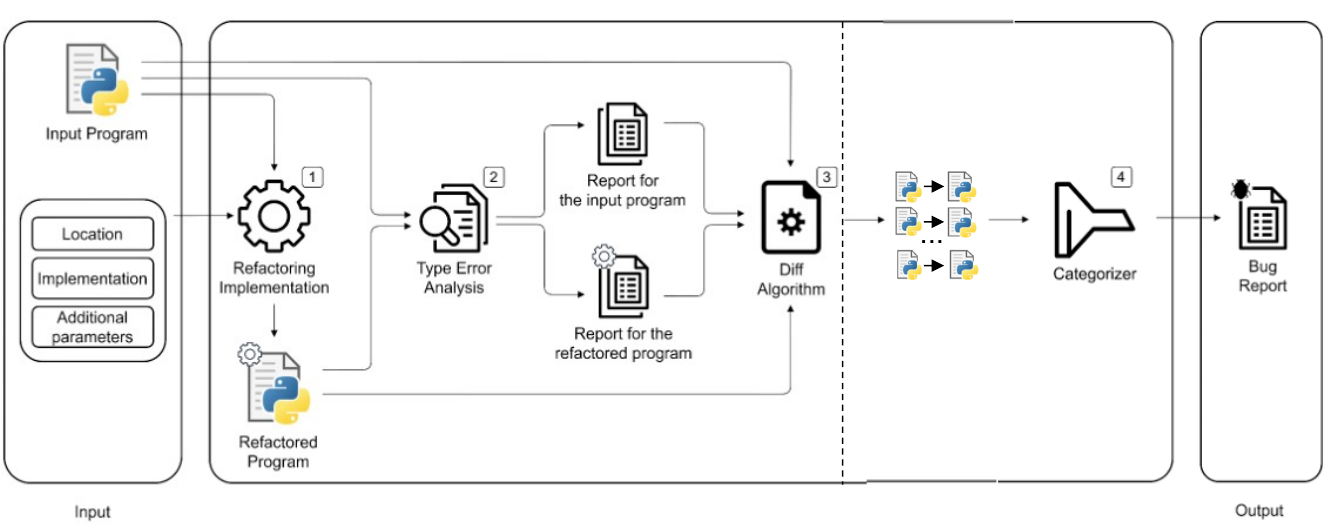}
    \caption{\review{A technique for detecting type errors in Python refactoring implementations.}}
    \Description{}
    \label{fig:technique}
\end{figure*}

Our technique takes as input a Python program, the refactoring implementation under test, the location where the refactoring should be applied, and any required parameters for the selected refactoring (Step~1).
The location parameter specifies the start and end offsets of the region in the source code to which the refactoring will be applied.
If the refactoring tool under test throws any exceptions during the transformation process, our technique records this outcome as-is in the generated bug report.
Next, we perform type error analysis on both the original and the refactored versions of the program (Step~2). 
The differential analysis algorithm then receives both versions of the program and their respective type-checking reports (Step~3). 
This algorithm identifies discrepancies and classifies the resulting failures (Step~4), ultimately assembling a bug report.
Figure~\ref{fig:technique} provides an overview of all steps in our technique.

\subsection{Apply Refactoring}
\label{subsec:techStep1}

In the first step, we apply the refactoring implementation.
For instance, consider the Rename Method refactoring presented in Section~\ref{sec:motivatingExample}. For this refactoring type, our technique receives Listing~\ref{lstMotivating:inputProgram} as the input program, the location of the \review{function}, and the new name to apply the refactoring (for example, \code{compare}).
The parameters consider the location as an element of the parameters to apply the refactoring instance. In this example, we use the Rope refactoring implementation.
Our algorithm identify the location of the \review{function} \code{\_\_lt\_\_} using Python's Abstract Syntax Tree (AST) library.
\review{Finally, it yields the refactored program.}
For other types of refactoring, we apply a similar strategy with adjusted parameters as needed.

\subsection{Type Error Analysis}
\label{subsec:techStep2}
\begin{figure*}[ht]
    \centering
    \begin{subfigure}{1\textwidth}
        \begin{subfigure}{\textwidth}
            \centering
            \includegraphics[keepaspectratio, width=1\textwidth]{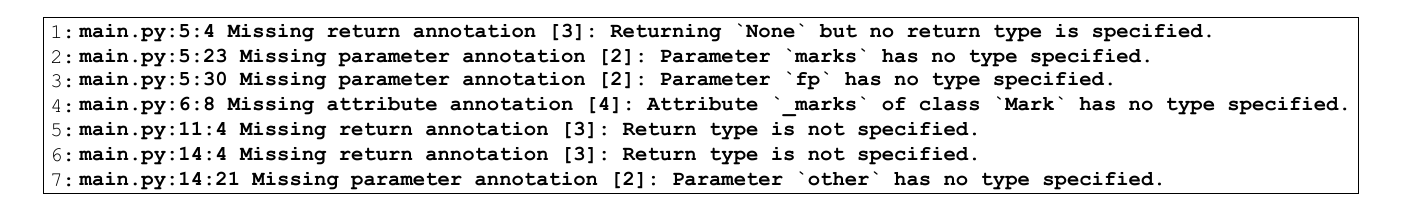}
            \caption{Type error report of the input program.}
            \label{fig:reportTypeError:report}
        \end{subfigure}
        \begin{subfigure}{\textwidth}
            \centering
            \includegraphics[keepaspectratio, width=1\textwidth]{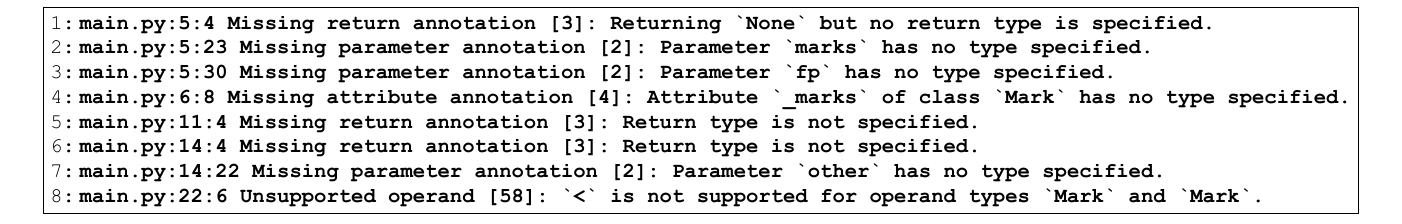}
            \caption{Type error report of the refactored program.}
            \label{fig:reportTypeError:reportOfRefactored}
        \end{subfigure}
    \end{subfigure}
    \begin{subfigure}{\textwidth}
    \centering
        \begin{subfigure}{.28\textwidth}
\begin{lstlisting}[numbers=none, frame=single, xleftmargin=0cm,  xrightmargin=0cm,  basicstyle=\ttfamily\scriptsize]
Missing return annotation [3]
Missing parameter annotation [2]
Missing parameter annotation [2]
Missing attribute annotation [4]
Missing return annotation [3]
Missing return annotation [3]
Missing parameter annotation [2]
\end{lstlisting}
            \caption{Type errors of the input program.}
            \label{reportTypeError:inputProgram}
        \end{subfigure}%
        \qquad
        \begin{subfigure}{.3\textwidth}
\begin{lstlisting}[numbers=none, frame=single, xleftmargin=0cm,  xrightmargin=0cm, framesep=1pt, basicstyle=\ttfamily\scriptsize]
Missing return annotation [3]
Missing parameter annotation [2]
Missing parameter annotation [2]
Missing attribute annotation [4]
Missing return annotation [3]
Missing return annotation [3]
Missing parameter annotation [2]
Unsupported operand [58]
\end{lstlisting}
            \caption{Type errors of the refactored program.}
            \label{reportTypeError:outputProgram}
        \end{subfigure}%
        \qquad
        \begin{subfigure}{.23\textwidth}
\begin{lstlisting}[numbers=none, frame=single, xleftmargin=0cm,  xrightmargin=0cm, aboveskip=1mm, basicstyle=\ttfamily\scriptsize]
Unsupported operand [58]
\end{lstlisting}
            \caption{Resulting set.}
            \label{diffOutputProgram}
        \end{subfigure}
    
    \end{subfigure}
    \caption{Type error analysis of the input and refactored programs.}
    \Description{}
    \label{fig:reportTypeError}
\end{figure*}

The technique executes the type error analysis on the input and refactored program versions (Step 2).
We instantiate our technique using \mbox{Pyre 0.9.18~\cite{pyre}} as our tool to detect type errors.
The tool analyzes one program at a time and produces one report for each program.
Consider the input program outlined in Listing~\ref{lstMotivating:inputProgram} and the refactored version.
The reports generated for input and refactored programs are presented in Figures~\ref{fig:reportTypeError:report} and~\ref{fig:reportTypeError:reportOfRefactored}, respectively.

\subsection{Differential Algorithm}
\label{subsec:techStep3}

In Step 3, our technique receives the reports generated from Step 2.
In this step, our goal is to identify the new type errors introduced by the refactored implementation in the resulting program.
For instance, consider the \mbox{Pyre 0.9.18~\cite{pyre}} report of the input program in Figure~\ref{fig:reportTypeError:report} and the report of the refactored program in Figure~\ref{fig:reportTypeError:reportOfRefactored}.
The type error can be identified between the file's name and the description of the detected type error.
\review{First, our algorithm removes the file name, line number, and all text following the colon, retaining only the core type error message, as illustrated in Figures~\ref{reportTypeError:inputProgram} and~\ref{reportTypeError:outputProgram}.}
Then, we focus only on the set of new type errors added after applying the refactoring.
We drew inspiration from the Differential Test~\cite{McKeeman-DTJ-1998} to identify differences in the reports.
The set of new type errors introduced in the refactored program is presented in Figure~\ref{diffOutputProgram}.

\subsection{Categorizer}
\label{subsec:techStep4}

\review{After applying all transformations from a refactoring implementation under test in Steps~1-3, some of them may introduce new type errors.
However, the resulting set of transformations can be large and diverse, making manual analysis both time-consuming and error-prone.
Additionally, multiple failures may stem from the same underlying bug.
To address these challenges, we developed a categorizer to group similar failures and streamline the analysis process.}

\review{We group the failures based on the type errors introduced after each transformation applied by a refactoring instance. These failure groups are then used to categorize the $N$ failures into bug candidates. For example, if $N$ \review{function} renamings result in the same type error as shown in Figure~\ref{diffOutputProgram}, we randomly select one representative from the group for manual analysis and consider it a potential bug candidate.}
As another example, during our evaluation of the Rename Method refactoring, we identified three transformations that introduced the following type errors: \typeError{Unsupported operand}, \typeError{Missing global annotation}, and \typeError{Incompatible variable type}. Additionally, one transformation resulted in \typeError{Call error} and \typeError{Invalid class instantiation}. For each group of similar type errors, we selected one representative transformation as a bug candidate.

\subsection{Bug report}
\label{subsec:bugs}

In the last step, we may have large input and refactored programs to be analyzed in each transformation.
Reporting bugs using large code snippets brings a complexity of understanding to developers.
So, we simplified the input program drawing inspiration from the delta debugging technique~\cite{pontes-fse-2019,Zeller-book-2009} to improve this scenario.
We remove some code snippets in the input program, apply the refactoring with the same parameters, and check whether the new refactored program yields the same new type error. Otherwise, we put back the removed code snippet. We repeat this process until we cannot remove any Python construct in the input program anymore.
For example, we utilized the code available on the TextBlob repository on GitHub\footnote{ \url{https://bit.ly/textBlobFile}} to reduce it to the example presented in Section~\ref{sec:motivatingExample} which represents the bug\footnotemark[\getrefnumber{motivateExampleIssue}] reported to developers.

\section{Evaluation}
\label{sec:evaluation}

In this section, we evaluate our technique.

\subsection{Definition}
\label{subsec:rqs}

The goal of this study~\cite{Basili1994} is to evaluate our technique for the purpose of analyzing the refactoring implementations for Python with respect to its ability to detect type errors from the viewpoint of researchers in the context of transformations applied to open-source projects. We address the following research questions:
\begin{description}
    \item[\textbf{RQ$_{1}$}] To what extent can our technique detect type errors in refactoring implementations? 
    To answer this question, we count the number of transformations applied by refactoring implementations that introduce type errors.
    \item[\textbf{RQ$_{2}$}] What are the type errors  detected by our technique? 
    To answer this question, we list the distinct type errors found by our technique.
    \item[\textbf{RQ$_{3}$}] To what extent are the reported bugs accepted by developers? 
    To answer this question, we count the number of bug reports accepted by developers.
\end{description}

\review{
\subsection{Methodology}
\label{subsec:planning}

\subsubsection{Subject}
We selected the \textit{TextBlob} project, version~0.17.1, as the subject of our evaluation. TextBlob is a natural-language processing library for Python (compatible with versions 2 and 3), consisting of over 3,000 lines of code. It has also been used in prior studies~\cite{dilhara-ICSE-2022,Tsantalis-ICSE-2018}.

\subsubsection{Refactoring Implementations}
We evaluated refactoring types supported by Rope version 1.3.0~\cite{rope}: Rename Field, Rename Method, Inline Method, Extract Method, Move Field/Method, and Use Function. These refactorings were selected based on their practical relevance and frequency of use in real-world development scenarios~\cite{Murphy-Hill-TSE-2012,golubev-fse-2021}. They also present challenges specific to Python's dynamic typing. Without static type checking, transformations such as renaming or moving methods and attributes may silently break code that depends on dynamic features like reflection, string-based patterns, or duck typing. Refactorings in Python often interact with constructs such as dynamically declared attributes (\code{__init__}), private methods using PEP 8 naming conventions, and operator overloading (e.g., \code{__lt__}, \code{__add__}). Moreover, method extraction and inlining can unintentionally eliminate necessary object context (\code{self}) or duplicate logic. Name shadowing is also a risk when renamed identifiers conflict with built-ins or global variables. 

\subsubsection{Tooling and Environment}
All experiments were performed on a 2.60GHz six-core machine with 32GB of RAM. We used Rope 1.3.0 to programmatically apply refactorings. Rope requires as input the root directory of the project, the file path containing the target, and the exact character offset of the target element. For type error detection, we used Pyre 0.9.18~\cite{pyre}, a static type checker developed and maintained by Meta.

\subsubsection{Procedure}
We first parsed the modules of the TextBlob project to extract all method and field names using the Python abstract syntax tree (AST). Then, for each refactoring type, we randomly selected a target name from the corresponding set of elements. Each execution of our technique involved a specific refactoring implementation, a selected location in the code, and the original source program. After applying the transformation, we analyzed both the original and the refactored versions using Pyre to detect newly introduced type errors. In our evaluation, we consider that a correct transformation should not introduce any new type errors. This automated setup allowed us to execute and analyze hundreds of transformations, which would not be feasible with a manually curated evaluation and increases the likelihood of uncovering bugs in refactoring implementations.
}

\subsection{Results}
\label{subsec:results}

\review{
\noindent\textbf{RQ$_1$.}  
Our technique applied a total of \totalTargets{} refactoring instances, identifying \totalFailures{} failures in four refactoring types. Among these, \totalCorrect{} transformations were correctly applied without introducing any type errors, while \totalCannotApply{} instances could not be executed due to limitations in the tools or internal crashes. In some cases, the tools themselves reported that the transformation could not be applied, such as when inlining functions with multiple return statements. 
Table~\ref{fig:tableDiscussion2} summarizes the number of available targets (\textit{Variables}), failures introducing new type errors, correct applications, and false positives, across different strategies and refactoring types. Strategies such as \textit{Rename Field – Method names} and \textit{Rename Method – Field names} demonstrated higher success rates, achieving 132 and 104 correct applications, respectively. Conversely, the \textit{Rename Field – Keywords} strategy exhibited the highest failure rate, with 166 failures out of 167 attempts. The \textit{Cannot Apply} column highlights cases in which the refactoring tools failed to produce output, typically due to tool crashes or unhandled edge cases.
}

\begin{table}[!htbp]
    \centering
    \resizebox{\columnwidth}{!}{%
    \normalsize
    \begin{tabular}{|lc|c|c|c|c|c|}
        \hline
        \rowcolor[HTML]{000000} 
        \multicolumn{1}{|c|}{\cellcolor[HTML]{000000}{\color[HTML]{FFFFFF} \textbf{Refactoring}}} & {\color[HTML]{FFFFFF} \textbf{Strategy}} & {\color[HTML]{FFFFFF} \textbf{\begin{tabular}[c]{@{}c@{}}Variables\\ (targets)\end{tabular}}} & {\color[HTML]{FFFFFF} \textbf{\begin{tabular}[c]{@{}c@{}}Cannot\\ Apply\end{tabular}}} & {\color[HTML]{FFFFFF} \textbf{Failures}} & {\color[HTML]{FFFFFF} \textbf{\begin{tabular}[c]{@{}c@{}}False\\ Positives\end{tabular}}} & {\color[HTML]{FFFFFF} \textbf{\begin{tabular}[c]{@{}c@{}}Correct\\ application\end{tabular}}} \\ \hline
        \multicolumn{1}{|l|}{Inline Method} & Method names & 150 & 50 & 21 & 15 & 64 \\ \hline
        \multicolumn{1}{|l|}{Rename Field} & Keywords & 167 & 1 & 166 & 0 & 0 \\ \hline
        \multicolumn{1}{|l|}{Rename Field} & Method names & 167 & 1 & 30 & 4 & 132 \\ \hline
        \multicolumn{1}{|l|}{Rename Method} & Keywords & 150 & 0 & 150 & 0 & 0 \\ \hline
        \multicolumn{1}{|l|}{Rename Method} & Field names & 150 & 1 & 39 & 6 & 104 \\ \hline
        \multicolumn{1}{|l|}{Use Function} & Method names & 28 & 6 & 0 & 0 & 22 \\ \hline
        \multicolumn{1}{|l|}{Extract Method} & Method names & 166 & 0 & 62 & 103 & 1 \\ \hline
        \multicolumn{1}{|l|}{Move Method/Field} & Method/Field names & 174 & 138 & 12 & 23 & 1 \\ \hline
    \end{tabular}%
    }
    \caption{\review{Summary of the automated application of refactoring implementations and selected strategies. 
    Strategy = Input types used as a parameter; 
    Variables (targets) = the number of available targets to apply the selected refactoring type; 
    Cannot Apply = the refactoring cannot be applied; 
    Failures = number of refactoring applications that yield new type errors; False Positives = number of false positives; 
    Correct Application = number of refactoring instances that do not introduce type errors.
    }}
    \label{fig:tableDiscussion2}
\end{table}

\review{
\noindent\textbf{RQ$_2$.}  
Our technique identified \totalDistinctTypeErrors{} distinct type errors. Table~\ref{fig:tableDiscussion} presents a summary of these errors, grouped by refactoring type and strategy used. The results reveal that certain type errors are recurrent across different refactoring implementations. The detected errors span several categories, including: (i) \textit{compilation issues} such as \typeError{Parsing failure} and \typeError{Unexpected keyword}; (ii) \textit{scope-related errors} like \typeError{Unbound name} and \typeError{Undefined attribute}; (iii) \textit{structural problems} such as \typeError{Call error} and \typeError{Unsupported operand}; and (iv) \textit{type violations} including \typeError{Inconsistent override} and \typeError{Invalid class instantiation}. These results demonstrate the diversity and depth of errors that our technique is capable of uncovering across multiple scenarios.
}
\begin{table*}
    \centering
    \includegraphics[keepaspectratio, width=0.8\textwidth]{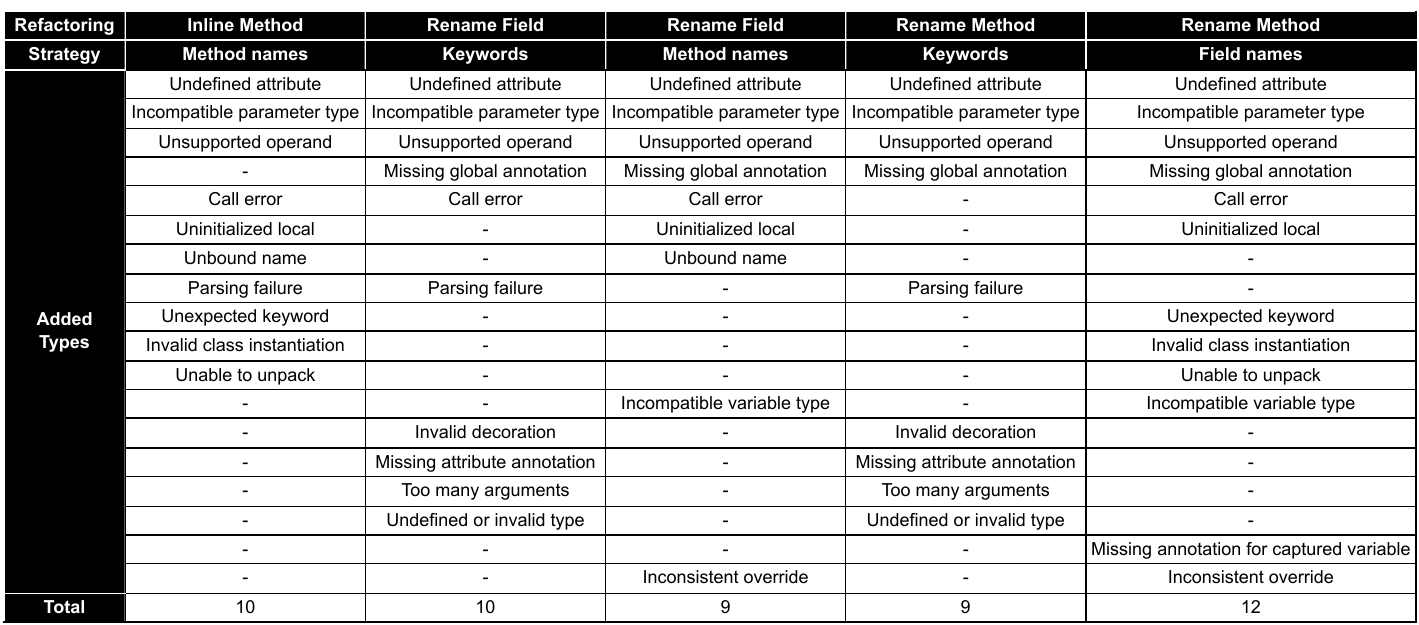}
    \caption{Summary of distinct detected type errors per refactoring type applied with Rope that emitted new type errors. 
    Strategy = parameters used by refactoring; Added Types = new type errors introduced by refactoring implementations.
    }
    \label{fig:tableDiscussion}
\end{table*}

\review{
\noindent\textbf{RQ$_3$.}  
After applying Step~4, we classified \totalFailures{} failures into 27 distinct bug reports. Table~\ref{fig:tableDiscussion3} provides a detailed overview of each reported bug. It includes both issues accepted by the JetBrains PyCharm team (IDs prefixed with `PY') and those submitted to the Rope repository on GitHub. In one case (ID 10), we reported an issue related to a specific refactoring type, but the fix was applied to two refactoring types; thus, we did not file a separate report. Additionally, Table~\ref{fig:tableDiscussion3} also includes manually identified bugs inspired by those automatically detected by our technique: one in JetBrains (ID 3) and another in the Rename Class refactoring (ID 29). These examples suggest that discovering a bug in one refactoring implementation can help developers reason about and identify missing preconditions in other implementations.
}

\begin{table*}[!ht]
\centering
    
    \resizebox{\textwidth}{!}{
    \begin{tabular}{|c|l|p{18cm}|c|}
\hline
\rowcolor[HTML]{000000} 
{\color[HTML]{FFFFFF} \textbf{ID}} & {\color[HTML]{FFFFFF} \textbf{Refactoring}} & \multicolumn{1}{c|}{\cellcolor[HTML]{000000}{\color[HTML]{FFFFFF} \textbf{Issue}}} & {\color[HTML]{FFFFFF} \textbf{Issue trackers IDs}} \\ \hline
1 & Inline Method & Inline Method refactoring is allowed in methods of the descriptor protocol & 742 \\ \hline
2 & Inline Method & Inline method refactoring inserts an unexpected argument & 757 \\ \hline
 & Inline Method & Cannot inline functions with the same name as different functions from another package used in the module &  \\ \cline{2-3}
\multirow{-2}{*}{3} & Rename Method & Refactoring custom methods touches library methods of the same name & \multirow{-2}{*}{PY-66251} \\ \hline
4 & Inline Method & Applying the Inline Method refactoring does not add the required import & 743 \\ \hline
5 & Inline Method & Inline Method refactoring is allowed in abstract methods & 744 \\ \hline
6 & Inline Method & Compilation error after applying the Inline method refactoring & 745 \\ \hline
7 & Inline Method & Inline method refactoring applied to rich comparison methods & 758 \\ \hline
8 & Inline Method & Inline method refactoring changes variables names after applying the transformation & 759 \\ \hline
9 & Inline Method & Inline method refactoring passes the wrong parameter to the inlined function body & 760 \\ \hline
 & Rename Method &  &  \\ \cline{2-2}
\multirow{-2}{*}{10} & Rename Field & \multirow{-2}{*}{Rename refactoring allow the use of Python keywords} & \multirow{-2}{*}{698} \\ \hline
11 & Rename Field & Rename Field refactoring allows you to rename a field with the same name used in a global method & 761 \\ \hline
12 & Rename Field & Rename Field refactoring allows you to use the name of special methods as a new name & 762 \\ \hline
13 & Rename Field & Rename refactoring doesn't rename a function's default arguments when the renamed variable is defined in the class scope & 686 \\ \hline
14 & Rename Field & Rename Field refactoring allows the use of declared method names as new field names & 763 \\ \hline
15 & Rename Field & Rename Field refactoring allows the use of a name that can not be iterable & 764 \\ \hline
16 & Rename Field & Rename Field refactoring allows you to rename a class field with class method names & 765 \\ \hline
17 & Rename Field & Rename Field refactoring allows you to rename a local field with the method name & 766 \\ \hline
18 & Rename Field & Rename Field refactoring allows you to change the method parameter, causing inconsistent override & 767 \\ \hline
19 & Rename Method & Rename Method refactoring allows the use of previously declared field name & 768 \\ \hline
20 & Rename Method & Rename Method refactoring allows you to rename a method to a name with an 'internal use' indicator & 769 \\ \hline
21 & Rename Method & Rename Method refactoring does not change the name of the super method in the classes that override it & 770 \\ \hline
22 & Rename Method & Rename Method refactoring is allowed in methods of the descriptor protocol & 771 \\ \hline
23 & Rename Method & Rename Method refactoring does not change the calls of the renamed method & 772 \\ \hline
24 & Rename Method & Rename Method refactoring allowed for methods designed for numeric type emulation & 773 \\ \hline
25 & Rename Method & Rename Method refactoring does not rename all implementations of an abstract method & 774 \\ \hline
26 & Rename Method & Rename Method refactoring allowed for methods defined to implement container objects & 775 \\ \hline
27 & Rename Method & Rename Method refactoring allows you to rename a nested method with a parameter name & 776 \\ \hline
28 & Rename Method & The Rename Method refactoring is allowed in the overridden method & 746 \\ \hline
29 & Rename Class & Rename refactoring doesn't apply to the references of the renamed class & 700 \\ \hline
\end{tabular}%

    }
    \caption{Summary of the bugs reported to Rope project. ID = identifier; Refactoring = refactoring type; Issue = short description of the reported bug; Issue trackers IDs = the ID of the reported issue in the project's repository (numbers starting with `PY'' represent JetBrains issue tracker (Link: \textit{{\small https://youtrack.jetbrains.com/issue/PY-<ID>}}); otherwise, the GitHub (Link: \textit{{\small https://github.com/python-rope/rope/issues/<ID>}}) repository).}
    \label{fig:tableDiscussion3}
\end{table*}

\subsection{Discussion}
\label{subsec:discussion}

Next we discuss our results.

\subsubsection{Parameters}
\label{subsec:discussion:locations}

Refactoring implementations typically require specific parameters to apply transformations correctly. 
As illustrated in Section~\ref{sec:motivatingExample}, we demonstrated the application of the Rename Method refactoring to the program shown in Listing~\ref{lstMotivating:inputProgram}. 
In this scenario, two key parameters must be provided: the location of the \review{function} to be renamed and the new name to assign.
Based on the program in Listing~\ref{lstMotivating:inputProgram}, our technique identifies all \review{functions} in the code and applies the refactoring to a selected subset of them. 
Similarly, other refactoring types evaluated in our study require analogous parameters, such as the location of a field, a new identifier name, or a \review{function} body to be extracted or inlined.

To guide the refactoring transformations in our technique, we employ three name selection strategies: ``\textit{Project Method Names}'', ``\textit{Project Field Names}'', and ``\textit{Keywords}''. 
These strategies are inspired by prior work~\cite{Schafer-OOPSLA-2010,Soares-TSE-2013,Schafer-PLPV-2009}, which used similar mechanisms to reveal bugs in refactoring implementations.
The ``\textit{Project Method Names}'' and ``\textit{Project Field Names}'' strategies extract existing \review{function} and field names from the target project, while the ``\textit{Keywords}'' strategy uses reserved keywords from the Python language as input values. 
For example, the bug presented in Section~\ref{sec:motivatingExample} was uncovered using the ``\textit{Project Method Names}'' strategy.

These strategies are particularly relevant in Python, where the language permits \review{functions} and fields to share names across global and local scopes. 
This flexibility introduces ambiguity, increasing the complexity of correctly applying refactorings -- particularly when deciding whether or not a transformation should be performed in a given context. 
Furthermore, using Python keywords as candidate names allows deeper exploration of corner cases and potential semantic conflicts within refactoring implementations.

Applying these strategies, we identified a total of \totalBugs{} bugs. 
The ``\textit{Keywords}'' strategy uncovered \totalKeywords{} bug in the Rename Field and Rename Method refactorings (both handled by a single Rope implementation). 
In addition, the ``\textit{Project Field Names}'' strategy led to \totalFields{} bugs (affecting Rename Method and Rename Class), while the ``\textit{Project Method Names}'' strategy revealed \totalMethods{} bugs across Rename Field, Rename Method, Inline Method, and Rename Class refactorings.
As future work, we plan to extend these strategies to incorporate additional naming patterns, such as special characters or combinations of alphanumeric and symbolic characters, to further challenge the robustness of refactoring implementations.

\subsubsection{Type Errors}
\label{subsec:discussion:errorsDetected}

Our technique using different strategies detected the \typeError{Undefined attribute}, \typeError{Incompatible parameter type} and \typeError{Unsupported operand} type errors in 3 refactoring types (Inline Method, Rename Field, and Rename Method). 
Pyre~\cite{pyre} emits \typeError{Undefined attribute} when an attribute (like a \review{function} or a property) is accessed on an object, and Pyre cannot find this attribute in the class definition or inferred type of the object.
For example, in Listing~\ref{lstpyre1} the \review{function} \code{show\_age} tries to print the \code{age} attribute of a \code{Dog} object.
However, the \code{age} attribute is not defined anywhere in the \code{Dog} class, leading to the \typeError{Undefined Attribute} type error when checked by Pyre.

The type error \typeError{Incompatible parameter type} is thrown when an argument into a function call does not match the expected parameter type of that function.
For instance, send a string to a function that expects an integer.
The type error \typeError{Unsupported operand} refers to operators not supported $-$ for example, \code{$a < b$} when \code{$a$} is a class that not accept this comparison type.
Both type errors are detected by our technique during the application of the Inline Method, Rename Field, and Rename Method refactoring types.

\begin{lstlisting}[language=Python, caption={An example demonstrating an \textit{Undefined Attribute} type error within the class definition itself.}, label=lstpyre1]
class Dog:
    def show_age(self):
        print(self.age) # 'Undefined Attribute'
\end{lstlisting}

\begin{lstlisting}[language=Python, caption={An example within a class construct demonstrating a \textit{Parsing Failure} type error.}, label=lstpyre2]
class Cat    # 'Parsing Failure'
    pass
\end{lstlisting}

Another type errors detected are the \typeError{Parsing failure} and \typeError{Unexpected keyword} type errors.
These type errors occur when the source code does not comply with Python's conventions.
For example, in Listing~\ref{lstpyre2} the error is due to the missing colon (:) at the end of the class declaration line (class \code{Cat}).
This is a syntax error, as Python expects a colon at the end of a class definition line.
This kind of mistake would prevent Pyre from parsing the class correctly, leading to the \typeError{Parsing Failure} error.
Our technique identified 3 refactoring types (Inline Method, Rename Field, and Rename Method) introducing those errors.

A list of distinct type errors identified after executing the technique for various refactoring implementations may be found in Table~\ref{fig:tableDiscussion}.
For example, using the \review{function} names strategy when evaluating the Inline Method refactoring, our technique identified 10 distinct type errors introduced.

\subsubsection{Test Input Programs}
\label{subsec:discussion:testInput}

In the first step of our approach, a Python program is provided as input. 
The technique is versatile and can be applied to a variety of Python projects. For our evaluation, we selected the open-source project TextBlob, which has also been used in prior studies~\cite{dilhara-ICSE-2022, Tsantalis-ICSE-2018}.
TextBlob includes 77\% of Python's language keywords, offering substantial coverage that enabled the detection of multiple refactoring-related bugs. 
However, it does not include certain keywords such as \code{async}, \code{await}, \code{del}, \code{with}, \code{nonlocal}, \code{global}, \code{finally}, and \code{yield}. 
These omissions may limit our ability to detect type errors associated with language constructs that rely on these keywords.
As future work, we plan to broaden our evaluation by incorporating a more diverse set of Python programs that exercise a wider range of language features.

For instance, consider the code snippet provided in Listing~\ref{lstpyreAsync}.
Based on the parameter value, this code may conditionally create an async task.
If the parameter is set to false, the async task is not created, and attempting to use the await command in such a scenario would result in a type error.
In this example, Pyre would emit the \typeError{Incompatible Awaitable Type} type error, and our technique would detect those type errors when they are introduced in the refactored program.

\begin{lstlisting}[language=Python, caption={An example demonstrating an \textit{Incompatible Awaitable Type} type error.}, label=lstpyreAsync, morekeywords={async, await}]
from asyncio import create_task, sleep
async def create_new_task(flag):
    if flag:
        task = create_task(sleep(1))
    else:
        task = None
    await task
\end{lstlisting}

The code presented in Listing~\ref{lstpyreGenerator} contains an async generator\footnote{\url{https://docs.python.org/3/library/typing.html\#typing.AsyncGenerator}} that allows asynchronous iteration.
It is defined using \code{async def} and contains \code{yield} statements.
Async generators are ideal when you need to produce values over time while still maintaining the responsiveness of your application.
In this example, Pyre would emit the \typeError{Incompatible Async Generator Return Type} type error, and our technique would detect those type errors when they are introduced in the refactored program.

\begin{samepage}
\begin{lstlisting}[language=Python, caption={An example demonstrating an \textit{Incompatible Async Generator Return Type} type error.}, label=lstpyreGenerator, morekeywords={async, await}]
from typing import AsyncGenerator
async def f() -> int:
    yield 0
async def g() -> AsyncGenerator[int, None]:
    if False:
        yield 1
\end{lstlisting}
\end{samepage}

\subsubsection{Bugs}
\label{subsec:discussion:bugsDetection}

\review{
We used a combination of GitHub labels, developer comments, and issue resolution status to assess bug acceptance. While GitHub automatically assigns the "bug" label when an issue is reported, Rope maintainers can review and update the label during triage as needed. If the label remains after review, we interpret it as implicit acceptance. In some cases, developers changed the bug tag to enhancement. Explicit rejections typically include the invalid label. Issues closed with an associated fix were considered both accepted and resolved.
}

The transformation (ID 24) \review{shown in Section~\ref{sec:motivatingExample}} presents a bug\footnotemark[\getrefnumber{motivateExampleIssue}] reported to developers of Rope project during application of Rename Method refactoring.
Our technique identified a failure category with one type error: the \typeError{Unsupported operand} type error.
Pyre emits the type error due to the default value in the argument not being renamed after applying the refactoring implementation.
Rope developers accepted and indicated they would not correct one of the reported bugs.
They understood that the reported bug ID 29 (see Table~\ref{fig:tableDiscussion3}) should be the user's responsibility to ensure that the target name made to apply the Rename Class.
Moreover, the developers indicated that they may add a \review{function} in Rope's front-end interface to check whether the destination name exists and is available.

In some cases, the refactoring implementation crashes.
For example, consider Listing~\ref{lstCannotApply} as an input program to apply the Inline Method refactoring to the \review{function} \code{get_string}.
The refactoring implementation of Rope crashes. Our technique found a number of crashes similar to this one.
\begin{lstlisting}[language=Python, caption={A crash in Rope caused by the Inline Method refactoring.}, label=lstCannotApply]
def bar():
    s = get_string(1)%*\label{lstCannotApply:target}*)
    print(s)
def get_string(num):
    if num == 1:
        return 'hello'
    return ''
\end{lstlisting}

\review{Some bugs (e.g., IDs 22 and 24) are specifically related to Python’s dynamic typing and would not typically occur in statically typed languages. These include issues involving Python-specific constructs such as abstract functions, modules, attribute declarations, and private methods that follow PEP naming conventions, as well as type emulation behaviors using operators like \code{<} and \code{+}. }

\subsubsection{False Positives}
\label{subsec:discussion:falsePositives}

\review{
To ensure that program context is preserved, we analyze complete programs during Steps~1 and~2. In Step~3, however, we rely on diffs to isolate changes and identify type errors introduced by refactoring. This diff-based strategy helps reduce false positives that might otherwise arise from running Pyre on the entire refactored program. Nonetheless, it also introduces the risk of false negatives—particularly when subtle issues are not captured in the diff or when a bug is indirectly related to a change. Additionally, some bugs may be missed or misclassified during the grouping process in Step~4.
While Pyre itself may produce false positives, our approach of focusing solely on differences helps narrow the analysis to errors that are more likely to be introduced by the transformation. We recommend using the technique iteratively: developers can address a subset of detected issues and re-run the analysis to refine the results and capture remaining issues.

In some cases, developers did not accept the reported bugs, which may indicate potential false positives or reflect differing interpretations of refactoring correctness. For instance, in one of the reported cases, the Rope developers argued that it is the user's responsibility to avoid naming conflicts. However, in Java refactoring implementations, such conflicts are typically detected and prevented by the tool itself, as they can cause compilation or semantic errors. From our perspective, similar preconditions should also be enforced by Python refactoring tools, and such cases should be treated as bugs.
}

\subsubsection{Pytype}

Pyre may face some challenges related to false positive and false negatives.
An alternative tool for identifying type errors in refactored programs is Pytype\footnote{\url{https://google.github.io/pytype/}}, a static type analyzer developed by Google. 
Pytype analyzes Python code to check and infer types without requiring type annotations, flags common type-related mistakes, supports linting, and can optionally enforce user-provided type annotations or generate them in standalone files.
Unlike Pyre, which may exhibit limitations when analyzing dynamically typed or annotation-free code, Pytype is specifically designed to perform well even in the absence of type hints. 
Incorporating Pytype into our workflow may help reduce false negatives -- i.e., type errors that are not detected by Pyre—thereby enhancing the overall robustness of our type error detection process.

\begin{lstlisting}[language=Python, caption={Input program to apply the Inline Method refactoring.}, label=lstFalsePositive:input]
def normalizer():
    return ''
class Word:
    word = normalizer()%*\label{lstFalsePositive:target}*)
    def singular(self, parser=word):%*\label{lstFalsePositive:target2}*)
        return parser.singular('')
\end{lstlisting}

Some type errors detected by Pytype and Pyre are similar, as both tools are designed to identify type-related issues in Python code. 
However, their internal error classification schemes and naming conventions may differ.
For instance, when applying our technique to the transformation from Listing~\ref{lstFalsePositive:input} to Listing~\ref{lstFalsePositive:output}, Pytype reports a \typeError{name-error}, which corresponds to a type error that Pyre also detects. 
This demonstrates that our technique is capable of identifying type issues consistently across different type analysis tools.
As future work, we plan to integrate Pytype into our technique by adapting Step~3 (Section~\ref{subsec:techStep3}) to evaluate its effectiveness in uncovering additional bugs that may be missed by Pyre, potentially increasing the coverage and reliability of our analysis.

\begin{lstlisting}[language=Python, caption={Program after apply the Inline Method refactoring.}, label=lstFalsePositive:output]
def normalizer():
    return ''
class Word:
    word = normalizer()
\end{lstlisting}

\subsubsection{Other Tools}
\label{subsec:discussion:otherTools}

As a feasibility study, we verified whether the same bugs occur in IDEs like PyCharm\footnote{\label{pycharmLink} \url{https://www.jetbrains.com/pt-br/pycharm/}} Community 2023.1, PyDev\footnote{\label{pydevSite} \url{https://www.pydev.org/}} 10.1.4, and \mbox{VSCode}\footnote{\label{vscodeSite} \url{https://code.visualstudio.com/docs/python/python-tutorial}} 1.81.0. \review{For each bug, we used the same initial code and parameters described in the corresponding bug report, manually applied the refactoring in the IDE, and analyzed the output using our technique.}
Our analysis revealed that some of the same bugs are also present in PyCharm (ID 3\footnote{\url{https://youtrack.jetbrains.com/issue/PY-66251}} in Table~\ref{fig:tableDiscussion3}) and in PyDev (IDs 10 and 13 in Table~\ref{fig:tableDiscussion3}).
\review{These examples suggest that identifying a bug in one refactoring implementation can help developers reason about and uncover missing preconditions in others.}
As future work, we intend to broaden our evaluation to include additional IDEs and refactoring engines. To enable automated testing in new environments, it is necessary to access the refactoring implementation interfaces and adapt Step~1 of our technique (Section~\ref{subsec:techStep1}) accordingly. Notably, the remaining steps of the technique do not require modification, demonstrating its adaptability and potential for integration with diverse refactoring tools.

\review{
\subsubsection{Implementation Effort}

Although our technique is general and extensible, working with Rope posed practical challenges due to its limited documentation, which assumes prior familiarity with its internal architecture. This resulted in a steep learning curve and required engineering effort. Programmatic use of Rope demands a detailed understanding of its API. For instance, applying the Rename Field refactoring to an attribute located on the third line of a file requires computing the exact character offset, accounting for all preceding characters and line breaks. Familiarity with the refactoring implementations may help developers integrate our technique more easily.

\subsubsection{Unit Tests}
Unit tests can detect certain refactoring issues, particularly those that result in observable behavioral changes. However, their effectiveness is limited by the availability and quality of test suites, as well as the developer's ability to identify which parts of the system are affected by the refactoring~\cite{DBLP:conf/icsm/RachatasumritK12}. Our static analysis technique complements unit testing by automatically analyzing the static type environment before/after refactoring operations. By leveraging tools such as Pyre, it can detect type-related violations introduced by transformations, even in the absence of test cases. While effective at identifying type errors, our approach doesn't capture all semantic behavior changes. Ideally, it would be integrated with dynamic techniques that automatically generate tests for the refactored entities to verify behavioral preservation~\cite{saferefactor-ieee,mongiovi-scp-2014}. We plan to address this in future work.

}

\subsection{Threats to Validity}
\label{subsec:threats}

One potential threat lies in the manual analysis of refactoring outcomes. 
\review{Step~4 was manually performed by the first author. After the initial bug reports were created, the second author independently reviewed them. In a few cases—particularly those involving subtle aspects of Python’s semantics—there were disagreements, which were resolved through discussion with a third author.} 
This process is inherently error-prone and may be influenced by individual bias or misinterpretation. 
However, all identified bugs discussed were submitted to tool maintainers, and \review{some of them} were confirmed and accepted, which lends credibility to the analysis.

Another threat relates to the accuracy of the type error detection process. 
Our technique relies on Pyre to identify type errors through static analysis. 
Although Pyre is a widely used and robust tool, it may produce false positives or false negatives, particularly due to Python's dynamic typing and Pyre's limitations in handling complex typing scenarios. 
To extract the type error locations, our technique processes Pyre's verbose textual output, which may also introduce parsing inaccuracies. 
Despite these limitations, Pyre provides a practical and consistent basis for evaluating type safety in Python, and our analysis strategy was carefully designed to minimize false positives. \review{However, further evaluation is necessary.}

Our evaluation was limited to a subset of popular refactoring types, including Rename Field, Rename Method, Inline Method, Extract Method, Move Field/Method, and Use Function. 
While these are widely used in practice~\cite{Murphy-Hill-TSE-2012}, our findings may not extend to more specialized or complex refactorings. 
Future work will expand the scope to cover additional refactoring types and scenarios.

We evaluated refactoring implementations primarily from the Rope library. 
Although Rope is commonly used in the Python ecosystem, evaluating implementations from additional tools and IDEs such as PyCharm and VSCode would strengthen the generalizability of our conclusions. Some of the bugs identified in Rope were also observed in PyCharm and PyDev.
As discussed in Section~\ref{subsec:discussion:otherTools}, our technique is designed to be adaptable to other refactoring engines and type-checking tools.

Although we used real-world open-source projects, including TextBlob, and selected projects with at least 70\% Python keyword coverage (a metric used in prior studies~\cite{dilhara-ICSE-2022, Tsantalis-ICSE-2018}), these projects may not fully represent the diversity of Python programs in the wild. 
Different codebases may exercise other language features or rely more heavily on dynamic constructs, which could impact the effectiveness of our technique.

\section{Related Work}
\label{sec:relatedWork}

Opdyke and Johnson~\cite{Opdyke-SOOPPA-1990,Opdyke-PHD-1992} coined the refactoring term, describing the process and identifying common refactorings.
Roberts~\cite{Roberts-PHD-1999} automated the basic refactorings proposed by Opdyke.
Later, Tokuda and Batory~\cite{Tokuda-ASE-2001} demonstrated that the preconditions proposed by Opdyke are not sufficient to guarantee behavior preservation after applying transformations.
Moreover, proving refactorings concerning formal semantics considering all language constructs constitutes a challenge~\cite{Schafer-PLPV-2009}.
AlOmar et al.~\cite{AlOmar-infsof-2021} performed a systematic mapping study on behavior preservation during software refactoring, providing a comprehensive overview of current practices, challenges, and research gaps in the field.

Daniel et al.~\cite{daniel-fse-07} proposed an approach for automatically testing refactoring implementations in Java. 
Their technique is built on \textsc{ASTGen}, a Java program generator, and relies on a set of programmatic oracles to assess the correctness of refactorings. They developed six oracles to evaluate the outputs of refactoring transformations and applied the technique to 42 refactoring implementations. 
Building on this work, Gligoric et al.~\cite{Gligoric-ICSE-2010} introduced \textsc{UDITA}, a Java-like language that extends \textsc{ASTGen} by supporting a hybrid approach to test input generation. 
Unlike \textsc{ASTGen}, which follows a purely generative style, \textsc{UDITA} allows developers to express constraints using a combination of filtering and generation strategies, making it both more expressive and efficient in producing relevant test programs.
In contrast, our work focuses on testing refactoring implementations in Python rather than Java. Instead of generating synthetic test inputs, we leverage real-world open-source Python projects as input programs -- similar to the methodology adopted by Gligoric et al.~\cite{Gligoric-ecoop-13} in their empirical testing of refactoring implementations for Java and C.

Steimann and Thies~\cite{Steimann-ecoop-2009} found that mainstream Java IDEs -- such as Eclipse, NetBeans, and IntelliJ -- exhibit flaws in preserving accessibility during refactorings. They identified scenarios in which the application of common refactorings, such as Pull Up Members, leads to unintended changes in program behavior, highlighting the challenges of ensuring semantic correctness in automated transformations.
Soares et al.~\cite{Soares-TSE-2013} proposed a technique for detecting behavioral changes and compilation errors introduced by refactoring implementations in Java. 
By applying their approach to 29 refactoring implementations using automatically generated input programs, they identified 57 bugs related to compilation errors and 63 related to behavioral changes.
Similarly, Mongiovi et al.~\cite{Mongiovi-TSE-2018} introduced a technique to detect bugs caused by overly strong preconditions in Java refactoring engines. 
Their method involves generating small Java programs and injecting them into refactoring implementations. 
They also systematically disable portions of the refactoring logic to identify which preconditions prevent the transformation from being applied, classifying them as potentially overly strong preconditions.
In our work, we identified some bugs related to \review{type} errors in Python refactoring implementations by applying transformations to real open-source projects, rather than relying on the generation of small synthetic programs.

Tempero et al.~\cite{Tempero-ACM-2017} conducted a large-scale survey involving 3,785 developers to investigate the barriers to applying refactorings in practice. 
Their findings indicate that refactoring decisions are often influenced by non-design considerations, and one of the key reasons cited for avoiding refactoring is inadequate tool support.
In contrast, our work aims to strengthen the reliability of refactoring tool implementations by statically detecting type errors introduced during transformations. 
By improving tool robustness, our technique may contribute to increasing developers' confidence in using automated refactoring tools.

Sch\"{a}fer~\cite{Schafer-fwrt-2012} discussed the issues and challenges associated with developing refactoring tools for dynamically typed languages. 
He highlighted the complexity of specifying and verifying preconditions in such environments, as well as the difficulty in determining whether access to module members is safe during transformations.
In this context, our technique can support tool developers by statically identifying type errors introduced by refactoring implementations in Python, thereby helping to address some of the challenges outlined by Sch\"{a}fer.

Wang et al.~\cite{wang-2024} conducted a comprehensive manual analysis of 518 bugs from three widely used refactoring engines in Java -- Eclipse, IntelliJ IDEA, and NetBeans -- identifying common root causes, bug symptoms, and characteristics of input programs that trigger faults. Their study yielded a set of actionable findings, which were used to derive guidelines for improving the detection and debugging of refactoring-related defects. 
Furthermore, their transferability analysis revealed 130 previously unknown bugs in the latest versions of these tools, underscoring the widespread and persistent nature of \review{bugs} in refactoring implementations.
In contrast, our work evaluates a static analysis technique for detecting type errors introduced by refactoring implementations in Python, using real-world open-source programs as input. While Wang et al.~\cite{wang-2024} focus on a broad characterization of refactoring bugs in statically typed languages and IDEs, our approach targets type-related issues specific to dynamically typed environments.

Dong et al.~\cite{dong-icse-2025} proposed a ChatGPT-based approach for testing refactoring engines, leveraging LLMs to automatically generate test programs aimed at uncovering defects. Their method constructs a feature library derived from existing bug reports and test cases, defines preconditions for each refactoring type, and employs predefined prompt templates to guide the generation process. The generated programs are then used in differential testing across multiple refactoring engines, with the results manually analyzed to identify defects. By evaluating seven refactoring types, the authors identified 115 bugs that led to compilation errors or behavioral changes in Java refactoring implementations. In contrast, our approach introduces an automated static analysis technique to test refactoring implementations for Python using real-world open-source projects. 
Rather than generating synthetic programs through LLMs -- an approach that may be constrained by limited context windows -- we directly apply refactorings to existing Python codebases, enabling a broader evaluation of tool behavior in practical settings.

Gheyi et al.~\cite{gheyi2025evaluatingeffectivenesssmalllanguage} evaluated the effectiveness of Small Language Models (SLMs) in detecting two categories of refactoring bugs: transformations that introduce compilation or behavioral errors (Type I), and transformations that are incorrectly blocked by IDEs despite being valid (Type II). They evaluated eight language models, including \textsc{Phi-4 14B} and \textsc{o3-mini-high}, using zero-shot prompting to analyze 100 refactoring bugs reported in Java and Python, collected from widely used IDEs such as Eclipse and NetBeans. They highlighted the low computational cost of SLMs, their ability to generalize across languages and refactoring types, and promising results for identifying Type I bugs. However, the models struggled to detect Type II bugs with up to 61 lines of code and often failed to provide accurate explanations, limiting their reliability in complex scenarios.
Our work introduced a static analysis technique for detecting type errors introduced by refactoring implementations applied to Python projects. 
Unlike previous approach, which primarily focus on transformations applied to small, isolated programs due to context window limitations, our technique is capable of analyzing refactorings in larger, real-world programs. However, unlike previous work, our technique does not detect behavioral changes introduced by refactorings.
Through our evaluation, our technique uncovered \totalBugs{} bugs in four refactoring types.

Dilhara et al.~\cite{Dilhara-fse-2024} proposed PyCraft, a framework that combines static and dynamic code analysis with large language model (LLM) capabilities to refine code transformations for Python. PyCraft generates diverse code variations along with corresponding test cases to ensure correctness. To assess its effectiveness, the authors submitted 86 transformed code instances across 44 pull requests to open-source projects, achieving an acceptance rate of 83\%. These results underscore the potential of integrating LLMs with traditional analysis techniques to support automated code transformation.
In contrast, our work focuses on statically detecting type errors introduced by refactoring implementations. Our technique could be integrated into frameworks like PyCraft to increase confidence that the transformations do not compromise type safety.

\section{Conclusions}
\label{sec:conclusion}

In this paper, we introduced a static analysis technique to identify type errors introduced by refactoring implementations in Python. 
By applying \totalTargets{} refactoring transformations to a real-world Python project, our technique uncovered \totalBugs{} bugs of four distinct refactoring types, as well as \totalDistinctTypeErrors{} unique type errors.
We submitted all identified bugs to the respective tool maintainers, and \review{a number of them} of them were acknowledged and accepted -- demonstrating the practical relevance and real-world applicability of our findings.

These results provide empirical evidence that current Python refactoring implementations -- including those integrated into widely used IDEs -- can introduce type-related bugs, potentially undermining the safety and reliability of automated code transformations. 
Our findings highlight the need for more rigorous validation mechanisms to ensure that refactorings preserve type correctness, particularly in dynamically typed languages like Python. By acting as an additional verification layer, our technique supports developers and tool maintainers in strengthening the robustness of refactoring workflows and mitigating the risk of subtle, hard-to-detect bugs. Although our evaluation focused specifically on refactorings, the technique is not limited to this context and can be applied to assess other types of code transformations in Python by verifying whether they introduce new type errors.

Looking ahead, we intend to broaden our evaluation to include additional IDEs and refactoring engines, as well as explore the use of PyType as an alternative to Pyre for type error detection. 
We also plan to expand our technique to cover a wider range of refactoring types and to automatically detect other categories of refactoring-related bugs, such as behavioral changes and overly strong preconditions~\cite{Schafer-OOPSLA-2010,Soares-TSE-2013,daniel-fse-07,Mongiovi-TSE-2018,Soares-ICSM-2011}. 
Moreover, we aim to investigate the complementary role of Large Language Models (LLMs) and agent-based systems in validating and even repairing refactorings, paving the way for more intelligent and semantically aware refactoring support in future development environments.

\review{
\section*{Acknowledgments}
We thank the anonymous reviewers for their valuable feedback. This work was partially supported by CNPq (403719/2024-0, 310313/2022-8, 404825/2023-0, 443393/2023-0, 312195/2021-4), FAPESQ-PB (268/2025).
}


\end{document}